\documentclass[onecolumn]{svjour3}          
\usepackage{graphicx}
\usepackage{adjustbox}
\usepackage{amsmath}
\usepackage{epstopdf}
\usepackage[graphicx]{realboxes}
\usepackage{rotating}
\usepackage{float}
\usepackage{latexsym}
\usepackage{amssymb}
\usepackage{array}
\usepackage{longtable}
\usepackage{dcolumn}
\usepackage{bm}

\begin{document}

\title{Uniform rate inflation in $f(T, \mathcal{T})$-gravity }

\author{Priyanka Mandal, Bikash Chandra Paul}

\institute{ B. C. Paul \\
             \email{bcpaul@nbu.ac.in}\at
              Department of Physics, University of North Bengal, Siliguri, Dist. : Darjeeling 734 013, West Bengal, India \\
             }

\date{Received: date / Accepted: date}

\maketitle

\vspace{0.5in}

\begin{abstract}

We  present  uniform rate inflation in a modified $f(T, \mathcal{T}) $ gravity. It is found that early inflation  can be realized even with a scalar field field with a quadratic potential and a negative cosmological constant. We construct  inflationary model   for the inflaton field without
slow-roll approximation.  The inflaton field rolling at a constant speed permits a universe with sufficient inflation which encompasses the present universe fairly well. The initial value of the scalar satisfies a lower limit which is sufficiently large  compared to the field required for chaotic inflation. The Planck
prediction for cosmological perturbations are estimated. It is found that the cosmological model is stable.

\end{abstract}

\section{. Introduction}
\label{intro}

In modern cosmology it is accepted that the present universe emerged from an inflationary phase in the past \cite{1}. The concept of inflation was introduced to resolve some of the issues the standard Bigbang cosmology  cropped up when probed early universe. In addition to its success inflation  gives rise to a causal theory of structure formation \cite{2}  which predicted some cosmological observations which subsequently successfully verified. 
 During inflation quantum fluctuations of the scalar field that generated are amplified to cosmological scales which are also responsible for primordial tensor and scalar perturbations
constrained by cosmological observations namely,  cosmic microwave background (CMB) \cite{3}.
Although the concept of inflation came up 45 years back a huge numbers of inflationary models came up in the literature in different theories of gravity,  it is not yet clear when and how the universe entered into this phase of expansion. However, defining a slow roll parameter ($\epsilon$) interms of the Hubble parameter and its derivative, and it can be ascertained that the inflation  can be terminated when $\epsilon<< 1$. In 1983, Linde \cite{linde} proposed a temperature independent inflationary model known as "Chaotic inflation" is found to be more acceptable. It can be realized even in an anisotropic universe \cite{bc} and consistent in  anisotropic Brane world model \cite{bc1}. Inflationary model can be realized either with a massive or self-interacting scalar field but initial value of the scalar field must satisfy a lower limit for massive field $\phi > 3 M_p$ and weakly interacting in case of self interacting field \cite{lin}.

\section{Inflation in General Relativity}

The field equation in general theory of relativity is 
\begin{equation}
\label{1}
    R_{\mu\nu}-\frac{1}{2}  g_{\mu\nu} R = 8 \pi G \; T_{\mu\nu}
\end{equation}
where $G$ is the gravitational constant and $ T_{\mu\nu}$ energy momentum tensor where the stress of the energy momentum tensor is $(\rho, -p,-p,-p)$. In the presence of homogeneous scalar field $\phi$ we obtain the following :

\begin{equation}
\label{2}
    \rho = \frac{1}{2} \dot{\phi}^2 + V(\phi), \; \; \; p= = \frac{1}{2} \dot{\phi}^2 - V(\phi).
\end{equation}
We consider a  RW-metric given by
\begin{equation}
\label{3}
    ds^2= -dt^2+ a^2(t) \left[\frac{dr^2}{1-k r^2}+r^2(d\theta^2+sin^2\theta d\phi^2)\right]
\end{equation}
Using eqs. (\ref{1})-(\ref{3}) we obtain the components of the field equations which are
\begin{equation}
\label{4}
    3\left(\frac{\dot{a}^2}{a^2} +\frac{k}{a^2}\right)= 8 \pi G  \left( \frac{1}{2} \dot{\phi}^2 + V(\phi)\right)
\end{equation}
\begin{equation}
\label{5}
    2 \frac{\ddot{a}}{a} + \left(\frac{\dot{a}^2}{a^2} +\frac{k}{a^2}\right)= - 8 \pi G  \left( \frac{1}{2} \dot{\phi}^2 - V(\phi)\right)
\end{equation}
The conservation equation is given by
\begin{equation}
\label{6}
\ddot{\phi} + 3 \frac{\dot{a}}{a} \dot{\phi} +\frac{dV}{d\phi}=0
\end{equation}

For a flat universe a uniform rate inflation  proposed with a potential \cite{5} which is given by
\begin{equation}
\label{7}
V(\phi) = \frac{3 \lambda^2 \phi^2}{4} - \frac{\lambda^2}{2}
\end{equation}
The potential is determined by a quadratic scalar field with a negative cosmological constant. For a uniform rate inflation. the equation of motion can be solved and found
\begin{equation}
\label{8}
\dot{\phi} = - \lambda
\end{equation}
On integartion one gets 
\begin{equation}
\label{9}
\phi = - \lambda \, t +\phi_o
\end{equation}
where $\phi_o$ is an integration constant. It was later implemented in a braneworld model \cite{6} and determined the cosmological constant in brane

. The motivation of the present work is to implement the uniform inflation in a modified gravity $f(T, \mathcal{T})$ with $\mathcal{T}$ torsion.

\section{Teleparallel gravity}

The action for modified Teleparallel gravity proposed by Harko \cite{8} is given by 
\begin{equation}
\label{10a}
    \mathcal{A}= \int d^4x \; e \left( \frac{1}{16 \pi G} ( T + f(T, \mathcal{T}))+ L_m \right)
\end{equation}
where $e= det e^A_{\mu}= \sqrt{-g}$ and $\mathcal{T}$ is the torsion. The field equation is obtained from the action by varying it w.r.t.  the vierbein which is 
\[
 (1+f_T) \left[ e^{-1}_{\mu} (e e^{\sigma}_A S^{\nu \mu}_{\sigma}) -   e^{\sigma}_A T^{\alpha}_{\rho \sigma} S^{\rho \nu}_{\alpha} \right]+ (f_{TT} \partial_{\alpha} T + f_{T \mathcal{T}} \partial_{\alpha} \mathcal{T}) e^{\sigma}_A S^{\nu \alpha}_{\sigma}+ e^{\rho}_{A} \left( \frac{f+T}{4}  \right) 
 \]   
\begin{equation}
    \label{10b}    
 \;\;\;\;\;\;   -\frac{f_{\mathcal{T}}}{2}   \left( e^{\sigma}_A T^{\nu}_{\sigma} + p e^{\nu}_A \right)= 4 \pi G e^{\sigma}_A T^{\nu}_{\sigma}          
\end{equation}
where $ f_{\mathcal{T}}= \frac{\partial f}{\partial \mathcal{T}}$ and $f_{T\mathcal{T}}= \frac{ \partial^2 f}{\partial T \partial \mathcal{T} } $.
For a FRW universe  with the vierbein ansatz : $ e^A_{\mu}= diagonal \, (1, a(t), a(t), a(t))$ is the vierbin,  we obtain the following field equations:
\begin{equation}
\label{10c}
3 H^2 = 8 \pi G \rho_m - \frac{1}{2} (f+ 12 H^2 f_T) + f_{\mathcal{T}} (\rho+p)  
\end{equation}

\begin{equation}
\label{10d}
\dot{H} = - 4 \pi G (\rho+p)   -\dot{H} (f_T -12 H^2 f_{TT}) -  (\dot{\rho} -3 \dot{p}) f_{T \mathcal{T}} -f_{\mathcal{T}} \left( \frac{\rho+p}{2} \right) 
\end{equation}
We use scalar field to represent  $\rho$ and $p$ as given in eq. (\ref{2}).

\section{ Cosmological model in  ($T, \mathcal{T}$)-modified gravity}

We consider the gravitational action in eq. (\ref{10a}) as
\begin{equation}
\label{10}
\mathcal{A}_!= \int f(T,  \mathcal{T}) \sqrt{-g} d^4x 
\end{equation}
where 
$f(T, \mathcal{T})= \alpha T^n \mathcal{T}$  where $\alpha$ and $n$ are arbitrary constants. The field eqs (\ref{10c}) -(\ref{10d}), for $n=0$ become
\begin{equation}
\label{11}
3 H^2 = \frac{1}{2} (8 \pi G + 3 \alpha) \dot{\phi}^2 +  (8 \pi G -   2 \alpha)  V(\phi),
\end{equation}

\begin{equation}
\label{12}
\frac{\ddot{a}}{a} = = \frac{8 \pi G }{3} \dot{\phi}^2 +  \frac{1}{3} (8 \pi G -   2 \alpha) V(\phi).
\end{equation}
Thus the time dervative of the Hubble parameter is given by 
\begin{equation}
\label{13}
\dot{H} = - \frac{1}{2} (8 \pi G + \alpha) \dot{\phi}^2 
\end{equation}
For a homogeneous scalar field potential $V(\phi)= \frac{3}{4} \lambda^2 \phi^2 -V_0$ we rewrite eq. (\ref{11}) as 
\begin{equation}
\label{14}
3 H^2 = \frac{1}{2} (8 \pi G + 3 \alpha) \dot{\phi}^2 +  (8 \pi G -   2 \alpha) \left( \frac{3}{4} \lambda^2 \phi^2 - V_0\right)
\end{equation}
Using $\dot{\phi}$ from eq. (\ref{9}) we determine $V_0$ from eq. (\ref{4}) which is
\begin{equation}
\label{15}
V_0= \frac{ 8\pi G + 3 \alpha}{2(8\pi G-2 \alpha)} \lambda^2.
\end{equation}
Now we define $\frac{\dot{a}}{a} = \dot{\eta}$ and the 
eq. (\ref{11}) yields 
\begin{equation}
\label{16}
\dot{\eta} = \sqrt{  \frac{8\pi G - 2 \alpha}{4}} \lambda \phi
\end{equation}
Integrating the above equation, we get
\begin{equation}
\label{17}
\eta =  \frac{\lambda \sqrt{(8\pi G - 2 \alpha)}}{2}  \phi_0  t - \frac{\lambda^2 \sqrt{(8\pi G - 2 \alpha)}}{4}  t^2 + \eta_0
\end{equation}
where $\eta_0$ is a constant. Making a choice of $\eta(t=0) =0$, we can eliminate $\eta_0=0$. Integrating once again we get the scale factor of the universe which is given by 
\begin{equation}
\label{18}
a= e^{\eta} =  e^{    \frac{\lambda \sqrt{(8\pi G - 2 \alpha)}}{2}  \phi_0  t - \frac{\lambda^2 \sqrt{(8\pi G - 2 \alpha)}}{4}  t^2}
\end{equation}
which yields
\begin{equation}
\label{19}
a=  e^{- \frac{\lambda^2 \sqrt{(8\pi G - 2 \alpha)}}{4}  \left(t-\phi_0\right)^2 + \frac{\sqrt{(8\pi G - 2 \alpha)}}{4} \phi_0^2}
\end{equation}
The second derivative of the scale factor can be expressed as 
\begin{equation}
\ddot{a} =  \frac{\lambda^2 \sqrt{(8\pi G - 2 \alpha)}}{2} \left(\frac{\sqrt{(8\pi G - 2 \alpha)}}{2} \phi^2 \;-1\right)e^{ \eta} 
\end{equation}
Thus acceleration results when 
$ \phi > \sqrt{ \frac{2}{\sqrt{(8\pi G - 2 \alpha)}}}$.
  This is the phase of inflation. The universe comes out of inflationary phase when $ \phi_e = \sqrt{\frac{2}{\sqrt{(8\pi G - 2 \alpha)}}}$.
It leaves the horizon with sufficient inflation for number of e-folding : $\triangle \eta=60$ for an estimated initial scalar field given by
\begin{equation}
\label{20}
\phi_i^2 >   \frac{242}{ \sqrt{8\pi G -2 \alpha}}.
\end{equation}
It is evident that uniform inflation may start at $\phi_i >> 242$ for $\alpha >> 4 \pi G$. Using eqs. (\ref{8}) and (\ref{16}), we get
\begin{equation}
    \label{20a}
   \frac{d\eta}{d \phi}= - \frac{\sqrt{8 \pi G-2\alpha}}{4}\; \phi
\end{equation}
The above relation is used to determine the primordial density (curvature) perturbation \cite{7}. Again we have
$\delta N= - \delta \eta = - d \eta $ and therefore, we get 
\begin{equation}
\label{21}
 \delta N =    - \delta \eta = - \frac{d\eta}{d\phi} \delta \phi = \frac{8\pi G - 2 \alpha}{8\pi} \lambda \phi^2
\end{equation}
where $\delta \phi = \frac{H}{2\pi} $ and $H=\dot{\eta}$ from eq. (\ref{16}).  The spectrum  $P_R$ is given by 
\begin{equation}
\label{22a}
P_R =   (\delta \eta)^2 =  \frac{(8\pi G - 2 \alpha)^2}{64\pi^2} \lambda^2 \phi^4
\end{equation}
Using CMB prediction $\zeta = 5 \times 10^{-5}$, we calculate  the  coupling parameter if the scalar field for sufficient inflation making use of the initial value of the scalar field  $\phi_i= \sqrt{\frac{242}{\sqrt{8\pi G-2 \alpha}}}$ and the relation $P_R= \zeta^2$  which yields
\begin{equation}
\label{22}
\lambda = \frac{2 \pi }{\sqrt{8 \pi G - 2 \alpha}} \times 10^{-4}.              \end{equation}
In Fig. (1), it is evident that the strength of coupling parameter increases as $\alpha > - \infty$. It is maximum when $\alpha \leq 4 \pi G$, and gradually it is also possible to get an inflationary universe with sufficient inflation when  the strength of interaction of the scalar field diminishes.

\begin{figure}[H]
\centering
\includegraphics[width=8.41cm,height= 5.5 cm]{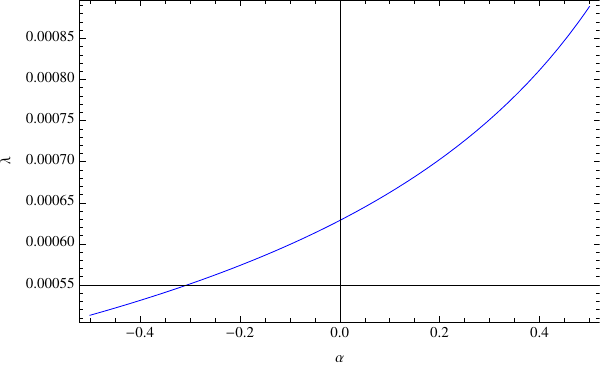}
\caption{ Variation of scalar field coupling constant ($\lambda$) with the coupling parameter of the modified gravity }
\label{Fig:1}
\end{figure}

The spectral index $n_s$ is 
\begin{equation}
    \label{23}
    n_s= 1+ \frac{d \ln P_R}{d \ln k } = 1+ \frac{d \ln P_R}{d \alpha} = 1 - \frac{8\pi G -2\alpha}{121}
\end{equation}
where $k$ is the comoving wave number of the perturbation. The spectrum for the tensor perturbation is
\begin{equation}
    \label{24}
    P_T=8 \left( \frac{H}{2\pi} \right)^2.
\end{equation}
Using eq;. (\ref{22a}) with $H (\phi)$ given by eq. (\ref{16}) we estimate the tensor to scalar perturbation is given by 
\begin{equation}
    \label{25}
    r=\frac{P_T}{P_R} = \frac{32}{\sqrt{8\pi G -2\alpha} } \; \frac{1}{\phi_i^2}    
\end{equation}

Using the Planck prediction  that an upper limit on tensor to scalar ratio $r < 0.044$  \cite{9}, we can estimate the limit on the coupling parameter $\alpha$.

\section{Discussion}

In the paper, we present uniform rate inflation in a modified  gravity $f(T, \mathcal{T})$ which can be realized with scalar field which is in te classical regime and with a very small strength of interaction. It is also evident from eq. (30) that the spectral index $n_S \rightarrow 1$ 
at the above limit and the observed tensor to scalar ratio can be used to estimate the gravitational coupling parameters. A weakly coupled scalar field is found to admit sufficient inflation in the modified gravity for $\alpha <0$. The stability of the model is studied in the section {\it Appendix: A}.\\

\section{Appendix A: Stability of the solution}

We discuss stability of the solution in this section: Consider a perturbation of the constant derivative scalar field $\dot{\phi}$ and that of the Hubble parameter as 
˙\[
\dot{\phi} \rightarrow \dot{\phi} + \epsilon,
\]
\[
\ddot{\phi} \rightarrow \ddot{\phi} + \dot{\epsilon}
\]
 \begin{equation}
 H= H + \delta H 
\end{equation}
we obtain 
\begin{equation}
    \delta H = \frac{(8 \pi G +3 \alpha)}{6 H )} \dot{\phi} \; \epsilon
\end{equation}
Using  eq. (\ref{8})  we obtain
\begin{equation}
\dot{\epsilon} + 3 H \epsilon=0
\end{equation}
Which can be integrated using $H$ and $\phi$ from eqs. (\ref{6}) and (\ref{16}), which is given by
\begin{equation}
    \epsilon = e^{- \frac{3 \sqrt{8 \pi G - 2 \alpha}}{2} \;  t (\phi_0-\lambda t)}.
\end{equation}
Taking  first order perturbation it is evident that 
as $\phi$ decreases  during
inflation, the perturbation in scalar field $\epsilon$  drops out very fast. Thus the solution obtained here is stable.\\

{\bf Acknowledgment :}
The authors (PM and BCP) would like to thank IUCAA, Pune and IUCAA Centre for Astronomy Research and Development (ICARD), NBU for extending research facilities. BCP would like to thank ANRF, Govt. of India  for a research grant.  \\


\end{document}